\documentclass[aps,prl,twocolumn,groupedaddress,showpacs,floatfix,superscriptaddress]{revtex4-1}
\usepackage{epsfig}
\usepackage{amsmath,amssymb}
\usepackage{graphicx}
\usepackage[dvipsnames,usenames]{color}
\usepackage[normalem]{ulem}
\usepackage{soul}  
\usepackage[colorlinks=true, citecolor = blue]{hyperref} 
\emergencystretch=\maxdimen
\begin{document}

\title{Quantum Monte Carlo Simulations of the 2D Su-Schrieffer-Heeger Model}    
\author{Bo Xing} 
\affiliation{Science, Mathematics and Technology Cluster, Singapore
  University of Technology and Design, 8 Somapah Road, 487372 Singapore}  
\author{Wei-Ting Chiu} 
\affiliation{Department of Physics, University of California, 
Davis, CA 95616, USA}
\author{Dario Poletti}
\affiliation{Science, Mathematics and Technology Cluster, Singapore
  University of Technology and Design, 8 Somapah Road, 487372 Singapore}
\affiliation{Engineering Product Development Pillar, Singapore
  University of Technology and Design, 8 Somapah Road, 487372 Singapore}
\affiliation{MajuLab, CNRS-UCA-SU-NUS-NTU International Joint Research
  Unit, 117542 Singapore}
\author{R.T. Scalettar}
\affiliation{Department of Physics, University of California, 
Davis, CA 95616, USA}
\author{George Batrouni}
\affiliation{Universit\'e C\^ote d'Azur, INPHYNI, CNRS, 0600 Nice,
France}
\affiliation{MajuLab, CNRS-UCA-SU-NUS-NTU International Joint Research
  Unit, 117542 Singapore}
\affiliation{Centre for Quantum Technologies, National University of
  Singapore, 2 Science Drive 3, 117542 Singapore}
\affiliation{Department of Physics, National University of Singapore, 2
  Science Drive 3, 117542 Singapore}
\affiliation{Beijing Computational Science Research Center, Beijing,
100193, China}

\begin{abstract}
Over the last several years, a new generation of quantum simulations
has greatly expanded our understanding of charge density wave phase
transitions in Hamiltonians with coupling between local phonon modes
and the on-site charge density.  A quite different, and interesting,
case is one in which the phonons live on the bonds, and hence modulate
the electron hopping.  This situation, described by the
Su-Schrieffer-Heeger (SSH) Hamiltonian, has so far only been studied
with quantum Monte Carlo in one dimension.  Here we present results
for the 2D SSH model, and show that a bond ordered wave (BOW)
insulator is present in the ground state at half-filling, and argue
that a critical value of the electron-phonon coupling is required for
its onset, in contradistinction with the 1D case where BOW exists for
any nonzero coupling.  We determine the precise nature of the bond
ordering pattern, which has hitherto been controversial, and the
critical transition temperature, which is associated with a
spontaneous breaking of ${\cal Z}_4$ symmetry.
\end{abstract}

\date{\today}

\maketitle

%%%%%%%%%%%%%%%%%%%%%%%%%%%%%%%%%%%%%%%%%%%%%%%%%%%%%%%%%%%%%%%%%%

\underbar{\bf Introduction:} The Su-Schrieffer-Heeger (SSH)
model\cite{su79}, where lattice vibrations (phonons) modulate the
ability of electrons to tunnel between neighboring sites ({\it i.e.}
the hopping parameter), was proposed more than four decades ago as a
description of the Peierls-charge density wave (CDW) phase transition
to an ordered insulating state, driven by the lowering of the
electronic kinetic energy.  The SSH model considered this transition
in the context of polyacetylene, but the instability has long been
known to occur experimentally in other quasi-1D systems, including
conjugated polymers\cite{keiss92}, organic charge transfer
salts\cite{ishiguro90}, MX salts\cite{toftlund84}, and
CuGeO$_3$\cite{hase93}.

Concurrently with exploring the metal-CDW insulator transition, the
SSH paper\cite{su79} already recognized the possibility of topological
excitations with fractional charge.  Over the last decade, tunable
cold atom systems have achieved real space superlattices
\cite{folling07,sebbystrabley06}, enabling the emulation of the SSH
Hamiltonian\cite{atala13,meier16,lohse16} as a simple realization of
1D ``BDI" class topological insulators\cite{qi11,schnyder08} ({\it
  i.e.} possessing spin rotation, time reversal, and particle hole
symmetries).  Understanding the underlying ordered phases and phase
transitions in two dimensions (thermal and quantum), as presented
here, lays the foundation towards studying the competition between
electron-phonon and electron-electron interactions and the possible
existence of topological phases.

Early numerical work on the SSH model in 1D addressed whether the
Peierls distortion survives the inclusion of fluctuations in the
phonon field, and indicated\cite{hirsch83} a difference in behavior
between the spinless and spinful SSH models, where the latter was
argued to be always ordered (albeit with a reduced order parameter as
the phonon frequency increases), and the former to have order-disorder transitions \cite{zheng88,mckenzie96, Weber2020}. 
Subsequent numerical and renormalization group studies\cite{su82,barford06,bakrim07,schmeltzer86,marchand10} refined
this understanding, but generally confirmed that lattice fluctuations
do not induce metallic behavior for spinful fermions.  In contrast,
the original suggestions\cite{hirsch83,bakrim07} that a metallic phase
is absent in the spinful Holstein model, have been overturned by
subsequent large-scale simulations\cite{clay05,fehske08,greitemann15}.
Even so, the precise value of the critical coupling, as well as a
quantitative description of the Luttinger liquid parameters of the metallic
phase remain open\cite{greitemann15,hohenadler12}. 
Polaron and bipolaron formation, along with condensation into superfluid states, has also been an area of considerable activity \cite{marchand10,Sous17,sous18}. 

These interesting and challenging 1D numerical studies of the SSH
model have been extended to the 2D Lieb lattice \cite{Li19}, but not to the single orbital square lattice geometry. 
One measure of the difficulty
of analogous higher dimensional studies is the controversy concerning
the optimal bond ordering patterns even when the lattice distortion is
frozen and not allowed to fluctuate at all, a free electron problem.
Tang and Hirsch\cite{tang88} argued that a ${\bf q}=(\pi,\pi)$ phonon,
polarized along the $x$-axis (or $y$-axis), provides the largest
energy gain when a displacement $\delta$ is introduced to the frozen
lattice.  Subsequent work\cite{ono00,chiba04} challenged this result,
making the claim that the optimal energy was achieved by a
superposition of a broad spectrum of lattice momenta, rather than
individual values at the borders of the Brillouin zone.

Here we present our exact quantum Monte Carlo (QMC) results for the 2D
SSH model, with full quantum dynamics of the phonons.  We performed
the simulations for lattice sizes $8\times 8$, $10\times 10$ and
$12\times 12$ with periodic boundary conditions. Larger sizes are not
practical with current algorithms. Our key conclusions are the
demonstration, at half filling, of a finite temperature phase
transition to an insulating bond ordered wave (BOW) phase, and a
quantitative determination of $T_c$ and the associated compressibility
gap.  Most importantly, we determine the nature of the BOW pattern,
thus resolving a long-standing
question\cite{tang88,ono00,chiba04}. Furthermore, we present numerical
evidence that, in the ground state, the electron-phonon coupling must
exceed a finite critical value for BOW to be established, unlike in
the Holstein model where the Peierls CDW phase is present for any
finite coupling on a square lattice.

\underbar{\bf Model and method:} We study the two-dimensional
``optical'' SSH model governed by the Hamiltonian,
\begin{eqnarray}
  \nonumber
  H&=&-t\sum_{\langle i,j \rangle, \sigma} (1-\lambda{\hat X}_{ij})({\hat
    c}^\dagger_{i\sigma}{\hat c}^{\phantom\dagger}_{j\sigma} + {\hat
    c}^\dagger_{j\sigma}{\hat c}^{\phantom\dagger}_{i\sigma}) -\mu
  \sum_{i,\sigma} {\hat n}_{i \sigma}\\
  &&+ \sum_{\langle i,j\rangle} \left [\frac{1}{2M}{\hat P}^2_{ij}
    +\frac{M}{2}\omega_0^2{\hat X}^2_{ij} \right ],
  \label{sshham}
\end{eqnarray}
where ${\hat c}^{\phantom\dagger}_{i\sigma}$ (${\hat
  c}^{\dagger}_{i\sigma}$) destroys (creates) an electron of spin
$\sigma=\uparrow$,$\downarrow$ on site $i$ and $\mu$ is the electron
chemical potential. The bond operators ${\hat X}_{ij}$ and ${\hat
  P}_{ij}$, connecting near neighbor sites $\langle ij\rangle$, are
the phonon displacement and momentum, $M$ is an effective mass, $\omega_0$ is an oscillation frequency, and $\lambda=(g/t)\sqrt{2M\omega_0/\hbar}$ is the electron-phonon coupling constant. In the following, we work in units for which $\hbar=t=M=1$ and we do our simulations with $\omega_0=1$. It was shown in 1D that this model gives the same results as the traditional ``acoustic'' SSH model\cite{weber15} where there is a coupling between the different phonon degrees of freedom.

At half filling, the 1D model is known to be in the BOW
phase\cite{weber15,fradkin83,sengupta03,barford06,bakrim15}, in which
the expectation value of the kinetic energy alternates with period
$\pi$ down the chain, for any $g>0$. To determine the nature of the
ground state phase diagram and the finite temperature transition in
2D, we use the exact determinant quantum Monte Carlo (DQMC)
method\cite{blankenbecler81,scalettar89,noack91}. Our main interest is
the half filled case ($\mu =0$); we calculate several quantities
needed to characterize the phase diagram: $\langle K_{x(y)}\rangle
\equiv \langle {\hat c}^\dagger_{i\sigma}{\hat c}_{i+{\hat x}({\hat
    y})\,\sigma} \rangle$, (the average kinetic energies (KE) in the
$x$ and $y$ directions), $\langle X_{x(y)}\rangle$ (the average phonon
displacements in the $x$ and $y$ directions), which indicate when
$x$-$y$ symmetry is broken. In addition, the KE bond-bond correlation
function, $G_{K_{x(y)}}(r)\equiv\langle K_{x(y)}(i) K_{x(y)}(i+r)
\rangle$, is calculated; its Fourier transform (the structure factor,
$S_{K_{x(y)}}(k_x,k_y)$) indicates the ordering vector of possible
long range BOW.  Equilibration of DQMC simulations of electron-phonon
Hamiltonians is known to be challenging\cite{noack91,batrouni19a}.
Data shown were typically obtained by averaging over ten independent
simulations, each using ${\cal O}(10^5)$ sweeps of the lattice before
making measurements, to ensure thermalization had occurred.  DQMC
simulations scale as $N^3 \beta$, where $N$ is the spatial lattice
size and $\beta$ is the inverse temperature.  At large $\beta$, $N
\sim 10^2$-$10^3$ are accessible\cite{varney09,batrouni19b}.  Because
of long equilibration and autocorrelation times in el-ph models,
studies are typically limited to the lower end of this range.

\begin{figure}
\includegraphics[width=1
  \columnwidth]{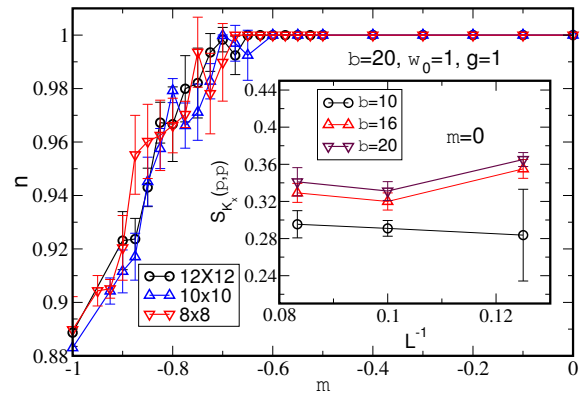}
\caption{(Color online) The density, $n$, versus the chemical
  potential, $\mu$. The gap is symmetric with respect to $\mu=0$ due
  to particle-hole symmetry. The inset shows that long range order,
  and therefore the gap, achieve their ground state values
for these lattice sizes for $\beta \gtrsim 16$.}
\label{rhovsmu}
\end{figure}

\underbar{\bf Results:} Figure \ref{rhovsmu} shows, for $g=1$, the
electron density $n$ versus $\mu$, clearly exhibiting a gapped phase
for which the compressibility $\kappa\propto \partial n / \partial
\mu=0$ in the regions $-0.7 \lesssim \mu \lesssim +0.7$.  The inset
shows $S_{K_x}(\pi,\pi)$ for several values of $L$ and $\beta$,
establishing that finite $\beta$ effects are negligible for $\beta\geq
16$. The structure factor $S$ provides a more exacting criterion for
convergence than a local observable like the density $n$ since it
involves long range correlations.

\begin{figure}
\includegraphics[width=1 \columnwidth]{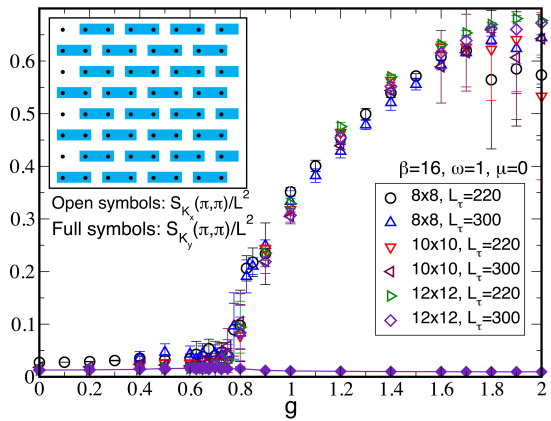}
\caption{(Color online) $x$ and $y$ kinetic energy bond-bond structure
  factors, $S_{K_x}(\pi,\pi)/L^2$ and $S_{K_y}(\pi,\pi)/L^2$, versus
  $g$. Long range checkerboard BOW develops for $g\gtrsim 0.75 \pm
  0.05$. $S_{K_y}(\pi,\pi)/L^2$ remains very small for all $g$
  indicating the order is only in the $x$ bonds.  Simulations shown
  here, and in subsequent figures, were begun in configurations
  favoring $x$ order.  Other starting points were tested to ensure
  results were independent of initial state.  See text for details.  }
\label{KESpipivsg}
\end{figure}

\begin{figure}
  \includegraphics[width=3cm]{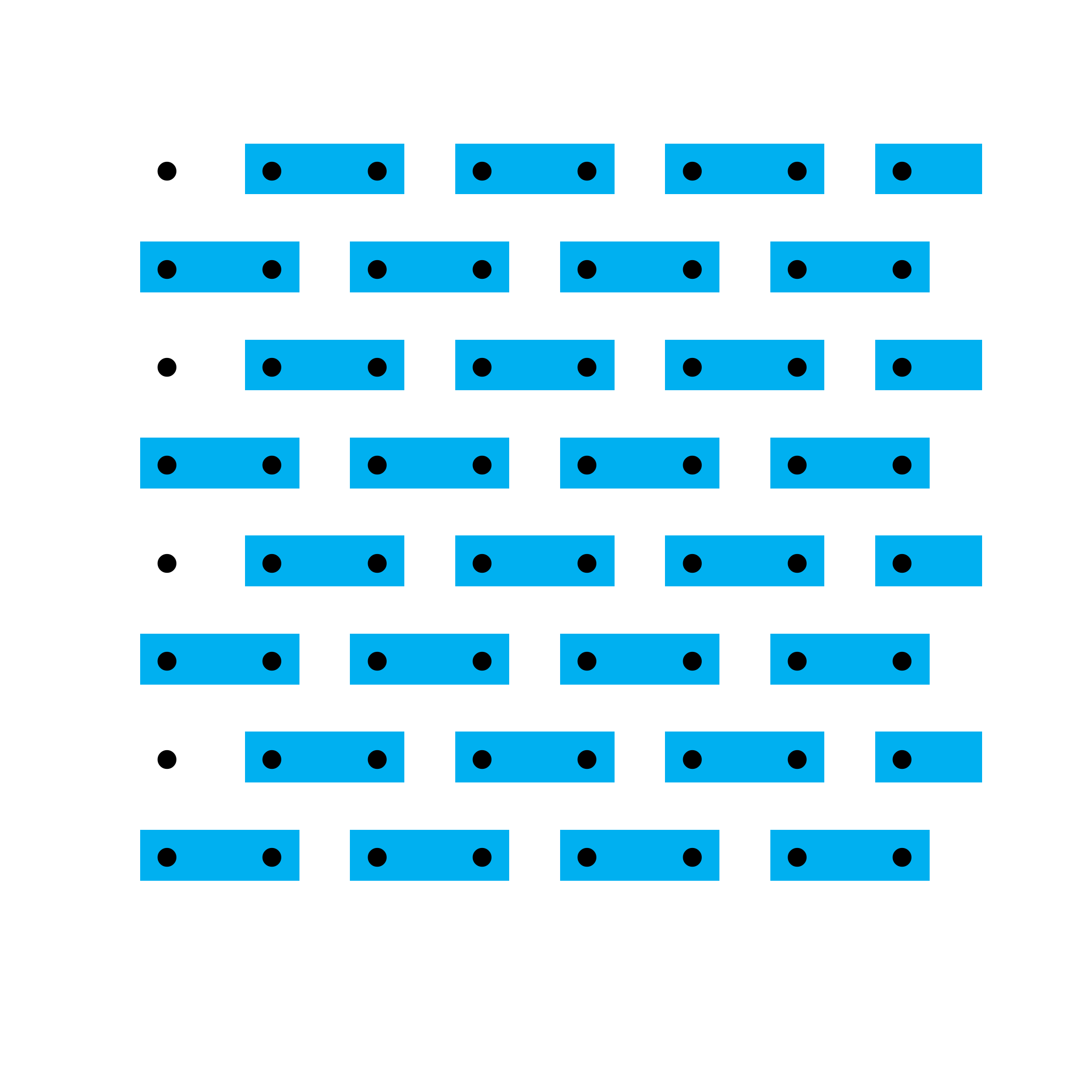}
  \includegraphics[width=3cm]{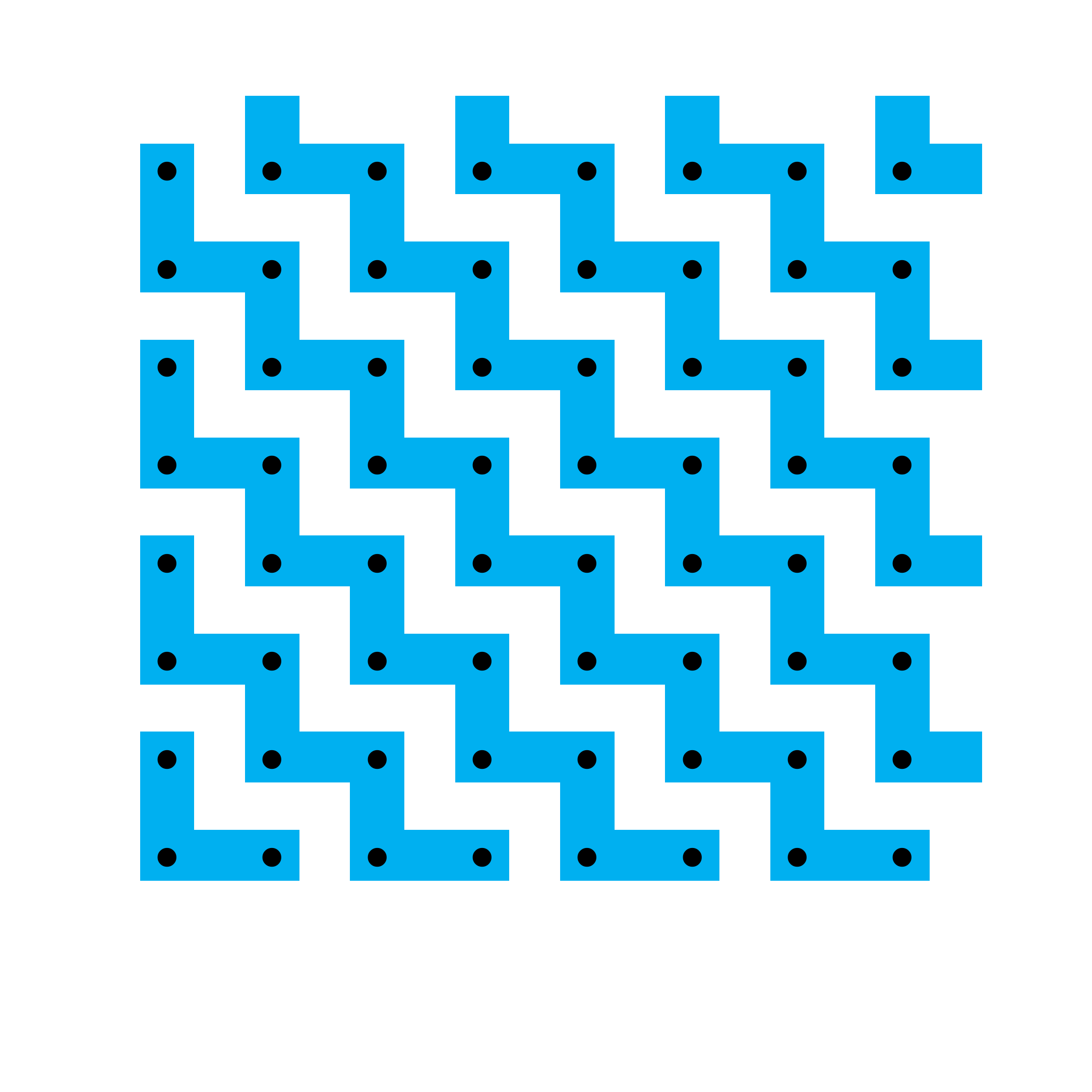}\\
\vskip -0.7cm
  \includegraphics[width=3cm]{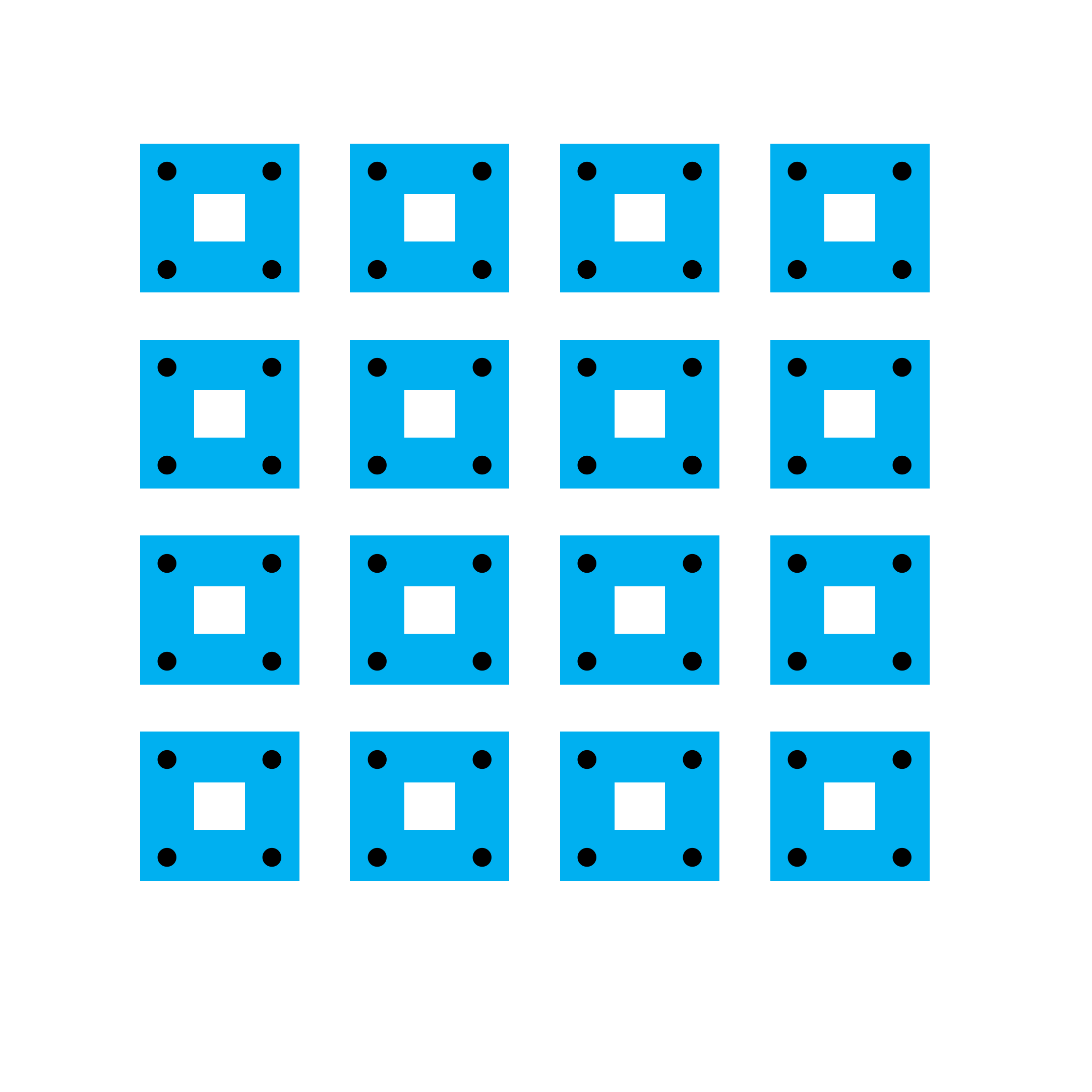}
    \includegraphics[width=3cm]{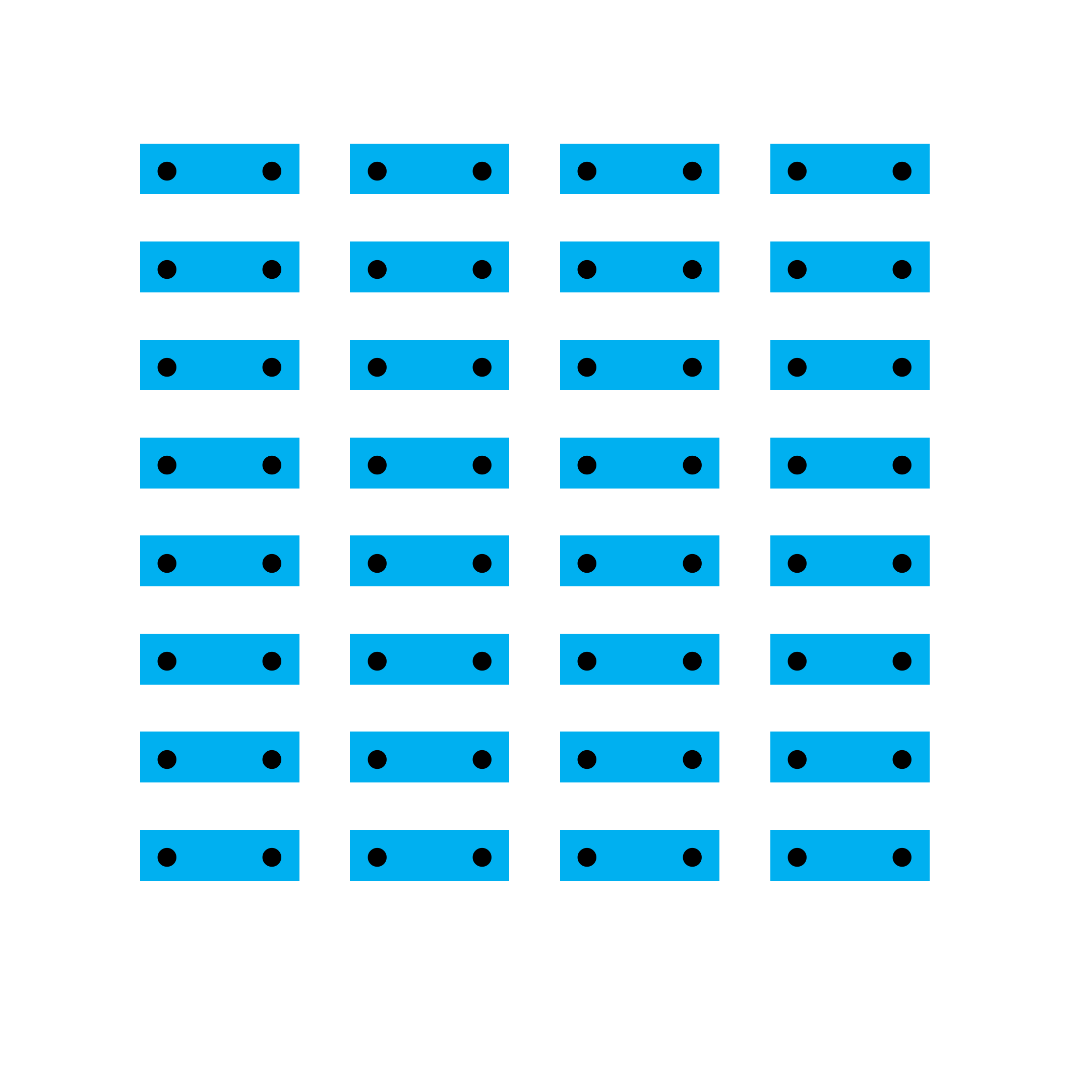}
\caption{(Color online) Possible BOW configurations. Top left:
  ($\pi,\pi$) order of $x$ bonds, top right: ($\pi,\pi$) order of $x$
  and $y$ bonds simultaneously, bottom left: ($\pi,\pi$) plaquette
  order, bottom right: columnar ($\pi,0$) of $x$ bonds. }
\label{initconfig}
\end{figure}

The nature of the gapped phase is exposed by studying
$G_{K_{x(y)}}(r)$ (or $G_{X_{x(y)}}(r)$) and the structure factor,
$S_{K_{x(y)}}(k_x,k_y)$. In Fig.~\ref{KESpipivsg} we plot
$S_{K_{x(y)}}(\pi,\pi)$ versus $g$.  At $g\approx 0.75 \pm 0.05$ a
quantum phase transition to BOW occurs where $S_{K_{x}}(\pi,\pi)$ or
$S_{K_{y}}(\pi,\pi)$ acquires nonzero value indicating symmetry
breaking. Starting simulations from several random or ordered phonon
initial configurations, Fig.~\ref{initconfig}, the system only develops
checkerboard BOW (for $g>g_c$) either for the $x$ or $y$ bonds but not
both simultaneously. This excludes all bond order patterns except the
one shown in the inset of Fig.~\ref{KESpipivsg} and the three
equivalent ones; the broken symmetry is ${\cal Z}_4$. To keep figures
uncluttered, we show BOW only for $x$ bonds.

\begin{figure}
\includegraphics[width=3.7in,height=2.8in]{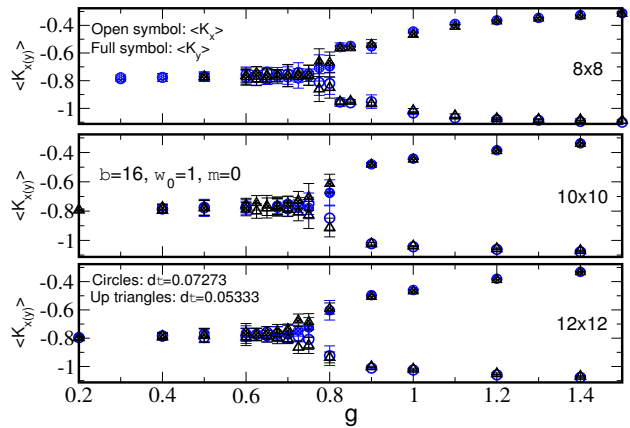}
\caption{(Color online) The average kinetic energies in the $x$ and
  $y$ directions, $\langle K_{x(y)}\rangle$, as functions of $g$ for
  three system sizes. The bifurcations in the average values indicate
  the phase transtion breaking the ${\cal Z}_4$ symmetry.  The inverse
  temperature $\beta=16$ ensures that the system is in its ground
  state for these spatial lattice sizes (see inset to
  Fig.~\ref{rhovsmu}). }
\label{Kxyvsg}
\end{figure}

\begin{figure}
\includegraphics[width=3.7in,height=2.8in]{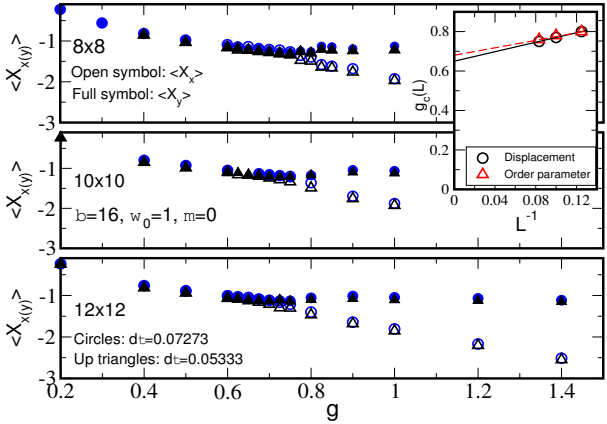}
\caption{(Color online) Same as Fig.~\ref{Kxyvsg} but for the phonon
  displacements.  The phonon displacement $\langle X_{ij}\rangle$
  becomes larger for $j=i+\hat x$, than for $j=i+\hat y$ at $g>g_c$,
  indicating a symmetry breaking quantum phase transition. Inset:
  Circles: extrapolation of the critical coupling from phonon
  displacement (this figure), $g_c=0.65$. Triangles: Extrapolation of
  critical coupling from the order parameter, Fig.\ref{Kxyvsg},
  $g_c=0.68$. }
\label{Xxyvsg}
\end{figure}

When a BOW forms, the $x$-$y$ symmetry breaking also manifests itself
in the $x(y)$ average kinetic energy, $\langle K_{x(y)}\rangle$, and
the average phonon displacements $\langle X_{x(y)}\rangle$. Figure
\ref{Kxyvsg} shows $\langle K_{x(y)}\rangle$ versus $g$ for systems
with $L=8,\,10,\,12$ and three values of the imaginary time step,
$d\tau\equiv \beta/L_\tau$ where $L_\tau$ is the number of imaginary
time slices in the DQMC. All $d\tau$ values give similar results
indicating that $d\tau$ Trotter errors are smaller than the
statistical error bars. Figure \ref{Kxyvsg} also shows that for $g
\gtrsim 0.75 $, the $x$-$y$ symmetry breaks with the formation of
$(\pi,\pi)$ BOW in the $x$ direction indicated by larger absolute
values for $\langle K_x\rangle$. These conclusions are supported by
Fig.~\ref{Xxyvsg} which shows the average $x$ and $y$ phonon
displacements for the same systems as in Fig.~\ref{Kxyvsg}. Again, the
$x$ and $y$ values bifurcate for $g\gtrsim 0.75$ signalling the
$x$-$y$ symmetry breaking and the formation of BOW. Extrapolating to
the thermodynamic limit the values of $g_c(L)$ from Figs.~\ref{Kxyvsg}
and \ref{Xxyvsg}, we find $g_c = 0.67 \pm 0.02$ (inset
Fig.\ref{Xxyvsg}). This relatively large finite value argues that,
unlike the one-dimensional SSH model which always displays BOW for any
finite $g$, in two dimensions a finite critical value of $g$ is needed
to establish BOW.

\begin{figure}
\includegraphics[width=1 \columnwidth]{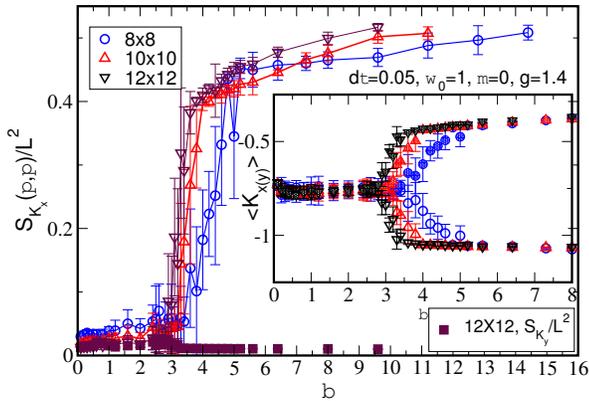}
\caption{(Color online) Main panel: BOW structure factor at
  ($\pi,\pi$) versus $\beta$ indicating a thermal phase
  transition. Inset: The average $x$ (open symbols) and $y$ (full
  symbols) kinetic energies as functions of $\beta$ showing a
  bifurcation consistent with the main panel. Finite size
  extrapolation gives $\beta_c\approx 1.9$ (see text).}
\label{SKxvsbeta}
\end{figure}

Next we study the transition from a disordered phase at high
temperature, $T$, to a BOW as $T$ is lowered. Figure \ref{SKxvsbeta}
shows $S_{K_x}(\pi,\pi)/L^2$ versus $\beta$ for system sizes
$L=8,10,12$. We see rapid increase in the structure factor as $\beta$
increases, indicating the establishment of a BOW. The transition
shifts to smaller $\beta$ (higher $T$) as $L$ increases. The inset
shows the bifurcation of $\langle K_{x(y)}\rangle$, and thus the
symmetry breaking, as $\beta$ is increased. We note the large error
bars in $S_{K_x}(\pi,\pi)/L^2$ in the transition region which are
caused by outliers which occur in about $10\%$ of the simulations.
\begin{figure}
\includegraphics[width=1 \columnwidth]{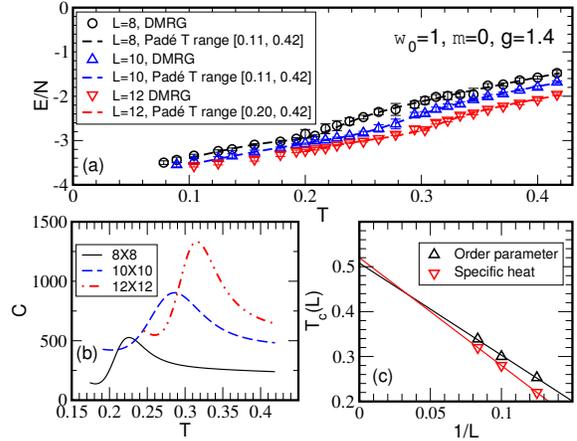}
\caption{(Color online) (a) The total energy per site as a function of
  the temperature, $T$, at half-filling. The lines are third order
  rational function (Pad\'e) fits to the data in the corresponding $T$
  intervals. (b) The specific heat, $C$ obtained for the derivatives
  of the Pad\'e fits. (c) Extrapolation to the thremodynamic limit of
  $T_C(L)$ obtained from the peaks of $C$ and from the order parameter
  (inset Fig.~\ref{SKxvsbeta}). The specific heat yields $T_C=0.52$ and
  the order parameter gives $T_c= 0.51$. }
\label{specificheat}
\end{figure}
We also study the finite $T$ transition by examining the specific
heat, $C={\rm d}E/{\rm d}T$, where $E$ is the total energy in the
ground state. We show in Fig.~\ref{specificheat}(a) $E/N$ as a function
of $T$ for three sizes, where $N$ is the number of lattice sites. The
lines through the symbols are obtained with a third order rational
function (Pad\'e) fit. In Fig.~\ref{specificheat}(b) we show $C$
obtained from the derivative of the Pad\'e fit, the positions of the
peaks agree very well with $T_c(L)$ obtained from the inset of
Fig.~\ref{SKxvsbeta}. Extrapolating to the thermodynamic limit $T_c(L)$
obtained from Fig.~\ref{SKxvsbeta} and from the peaks of $C$ yields
$T_c=0.51$ ($\beta_c=1.96$) and $T_c=0.52$ ($\beta_c=1.92$)
respectively, as shown in Fig.~\ref{specificheat}(c).

\underbar{\bf Conclusions:} In the past several years, a second
generation of QMC has been applied to Hamiltonians coupling phonon
modes to the local electron charge density. Critical temperatures and
quantum phase transition points for the square and honeycomb Holstein
models have been obtained to good
accuracy\cite{weber17,chen19,zhang19,cohenstead19}, building on
initial work which established the qualitative
physics\cite{scalettar89,noack91,vekic92,niyaz93,marsiglio90,hohenadler04}.

In this paper, we have reported the first results of QMC simulations
of the two-dimensional single orbital square lattice SSH model including full quantum dynamics, the
paradigmatic Hamiltonian describing phonons coupled to electron
hopping.  Previously, these had been undertaken only in one dimension
where they showed the system to be always in the BOW phase for any
finite value of the coupling. The ``obvious'' strong-weak bond
alternation pattern in 1D has a multitude of possible generalizations
in 2D, including staircase, columnar, staggered, and plaquette
arrangements\cite{hirsch89}, Fig.~\ref{initconfig}.  We have shown
here that in the ground state, a ${\bf q}=(\pi,\pi)$ order is
established for $x$ or $y$ bonds (but not both simultaneously). We
also showed that this bond ordering is accompanied by the opening of a
compressibility gap, given by a plateau in $\rho(\mu)$. Furthermore,
we have exposed an important qualitative difference between the two-
and one-dimensional SSH models: In 2D, a critical value of the
coupling, $g_c$, is necessary to trigger the quantum phase transition
from a disordered phase to the BOW, in contradistinction with 1D where
the BOW is present for any finite $g$ no matter how small
\cite{fradkin83,sengupta03,barford06,bakrim07,weber15}.

In the last year a variant of the 2D SSH model has been realized in a
number of new contexts, including acoustic networks\cite{zheng19} and
RF circuits\cite{liu19}.  These experiments allow for the observation
of edge states and associated topological invariants \cite{obana19}
within the context of the ``plaquette'' bond ordering pattern (bottom
left, Fig.~\ref{initconfig}). While this configuration can be
engineered artificially, our work shows that the low energy ordering
pattern which spontaneously arises from the simplest 2D SSH
Hamiltonian, Eq.~\ref{sshham}, consists instead of a staggered array
of dimers.  An interesting area of investigation will be modifications
to Eq.~\ref{sshham}, for example to the hopping parameters, which
might lead to the alternate ordering patterns of
Fig.~\ref{initconfig}, including plaquette arrangements.

\underbar{\bf Acknowledgments:} The work of RTS was supported by the
grant DE-SC0014671 funded by the U.S. Department of Energy, Office of
Science. We thank B. Cohen-Stead and S. Cui for helpful
discussions. D. P. acknowledges support from the Singapore Ministry of
Education, Singapore Academic Research Fund Tier-II (Project
MOE2016-T2-1-065). The computational work for this article was
performed on resources of the National Supercomputing Centre,
Singapore (NSCC)\cite{NSCC}.

%%%%%%%%%%%%%%%%%%%%%%%%%%%%%%%%%%%%%%%%%%%%%%%%%%%%%%%%%%%%%%%%%%%%%%%%%%%%
%%%%%     BIBLIOGRAPHY
%%%%%%%%%%%%%%%%%%%%%%%%%%%%%%%%%%%%%%%%%%%%%%%%%%%%%%%%%%%%%%%%%%%%%%%%%%%%

\bibliography{ssh}

\end{document}